\pdfoutput=1
\documentclass[aps,showpacs,pra,preprint,epsf,amsmath ]{revtex4}
\usepackage{graphicx}
\usepackage{bm}
\usepackage{amsmath}
\begin{document}
\sloppy
\title
{Lattice bosons in a quasi-disordered environment: The effects 
of a superlattice potential on single particle and many particle
properties}
\author{R. Ramakumar$^{1}$, A. N. Das$^{2}$, and S. Sil$^{3}$}
\affiliation{$^{1}$Department of Physics and Astrophysics, 
University of Delhi,Delhi-110007, India} 
\address{$^{2}$Saha Institute of Nuclear Physics,
1/AF Bidhannagar, Kolkata-700064, India} 
\address{$^{3}$Department of Physics, Visva Bharati,
Santiniketan-731235, India}
\date{15 April 2015}
\begin{abstract}
In this paper we present a theoretical investigation of the effect
of a superlattice potential on some 
properties of non-interacting bosons in one dimensional lattices 
with Aubry-And\'{r}e disorder potential. In the first part, we 
investigate the single particle localization properties. 
We find a re-entrant localization-delocalization
transition and the development of multiple mobility edges 
for a range of superlattice potential strengths.
In the second part, we study the Bose-Einstein
condensation with an additional harmonic trapping potential.
We find that an increase in the superlattice potential leads to
an increase in the depletion of the condensate in the low
temperature limit.
\end{abstract}
\maketitle
\section{Introduction}
\label{sec1}
 Studies of localization of the de Broglie waves
in disordered and quasi-disordered
environments have received renewed attention in the recent years
primarily due to the direct
observation of Anderson localization\cite{anderson} of these matter waves 
in cold atom experiments\cite{billy,kondov,jendrzejewski}. 
While random disorder localizes all states in one dimension\cite{mott}, 
a deterministic disorder distribution like the one in the Aubry-And\'{r}e
model(AA)\cite{aubryandre} leaves all states extended 
if the disorder strength
remains below a critical value. The AA model\cite{aubryandre} 
(or the Harper model\cite{harper}) 
has already been
experimentally realized 
and a detailed study of its localization properties 
conducted\cite{roati,lahini}.
In the AA model, since all the states become localized beyond 
a critical disorder strength, there is no mobility edge. 
Recent theoretical studies\cite{boers,biddle1,biddle2,biddle3,ribeiro,rds}
have discovered the development of mobility edges in several  
extended AA models. Continuing along this direction of research, in 
this paper we consider the effects of
a superlattice potential on the single particle and collective properties
of non-interacting bosons in a one dimensional AA model. 
Among other results, we find
a re-entrant single particle localization-delocalization transition 
with increasing superlattice potential strength.  
\par
The rest of this paper is organized as follows. The 
studies of the single particle localization properties 
are presented in Sec. II. 
The Sec. III deals with the effects of the superlattice potential on the
the Bose-Einstein condensation.
The conclusions are given in Sec. IV.
\section{The changes in the localization properties due to the
superlattice potential}
\label{sect2}
In this section, we consider the effect of a superlattice potential
on the single particle localization properties. Consider a lattice
boson moving in a one-dimensional lattice with AA disorder  and an
additional superlattice potential. For this system, the Hamiltonian is:
\begin{equation}
H=-t\sum_{<ij>}\left(c^{\dag}_{i}c_{j}+c^{\dag}_{j}c_{i}\right)
  +\sum_{i}\left[\lambda\,cos(2\pi qi)+(-1)^{i}V\right]c^{\dag}_{i}c_{i},
\end{equation}
where $t$ ($> 0$) is the the energy gain when a boson hops from 
site $i$ to its nearest neighbor (NN) site $j$, 
$c^{\dag}_{i}$ is a creation operator of a boson at site $i$,
$\lambda$ the strength of the AA potential, 
$q$ = ($\sqrt{5}$+1)/2 is the incommensurability parameter,
and $V$ is the strength of the superlattice potential.
Here $t$, $\lambda$, and $V$ have energy units.
All the energies in this paper are measured in units of $t$
and all the lengths are measured in units of the lattice parameter.
When $V$ = 0, one obtains the AA model and 
energy independent localization of all the states for $\lambda > 2$,
and thus there is no mobility edge. To study the effects
of finite $V$ on the localization properties, we first
write $H$ in the single particle site basis and then
numerically diagonalize it to obtain its eigen-energies and
eigen-functions. All our results presented in this paper are
\noindent for a lattice of 610 sites. We monitor the localization
of a given eigen-state with amplitude $a_i$ at site $i$
\begin{equation}
\displaystyle
\left|\psi\right> = \sum_{i}a_{i}\left|i\right>
\end{equation}
by calculating its Inverse Participation Ratio (IPR) defined by
\begin{equation}
\displaystyle
IPR =\frac{\sum_{i}p_{i}^{2}}{\left(\sum_{i}p_{i}\right)^{2}},
\end{equation}
where,
\begin{equation}
p_{i}=\left|<i|\psi>\right|^{2}=|a_{i}|^{2}.
\end{equation}
\begin{figure}
\begin{center}
\includegraphics[angle=270.0,width=4.0in,=4.0in]{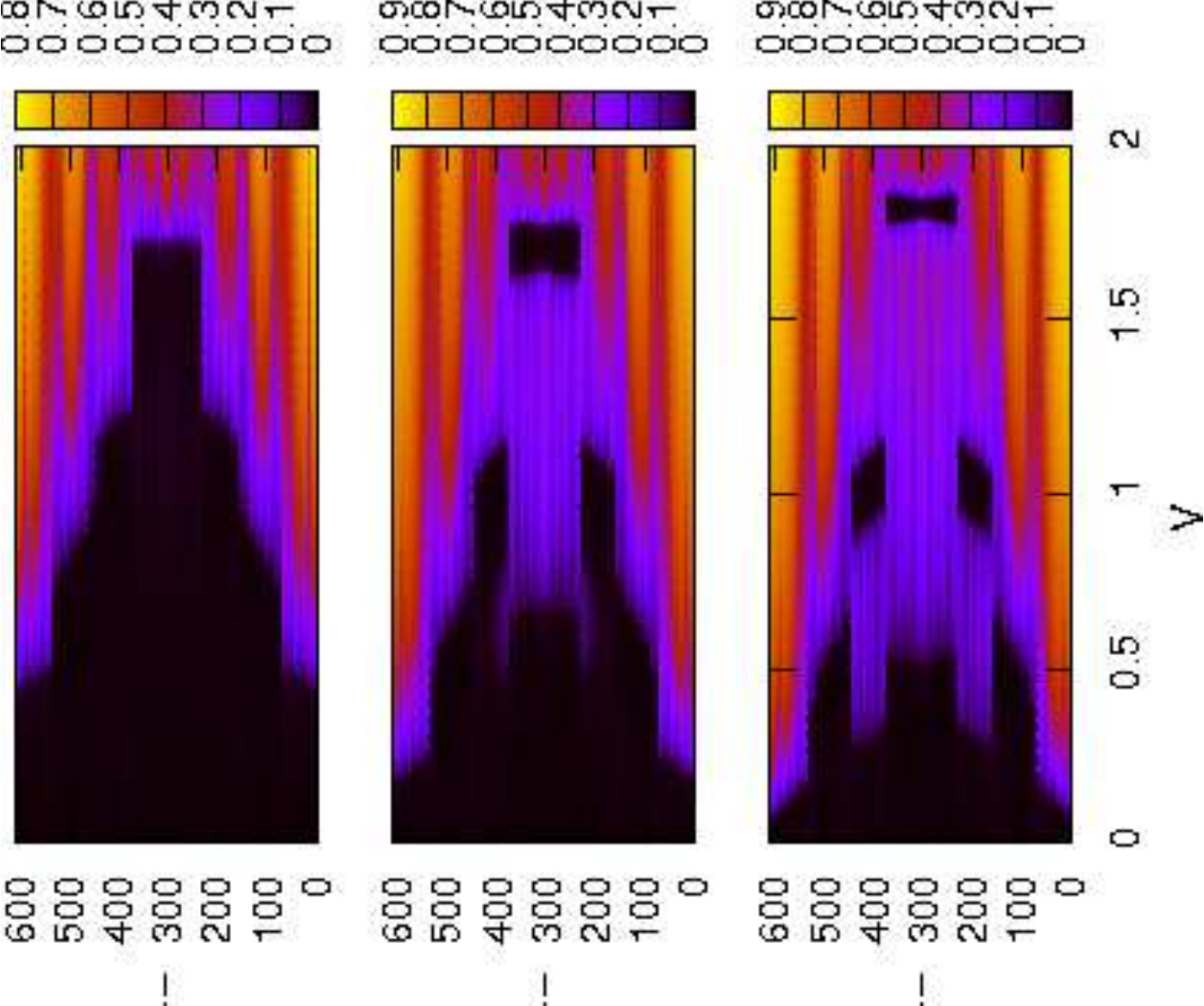}
\caption{  
The IPR values (in color) for different energy eigen-states as
a function of V for $\lambda$  = 1.0 (top panel), 1.3 (middle),
and 1.5 (bottom) for a closed chain of 610 sites.
}
\label{fig1}
\end{center}
\end{figure}
\begin{figure}
\begin{center}
\includegraphics[angle=270.0,width=4in,=4in]{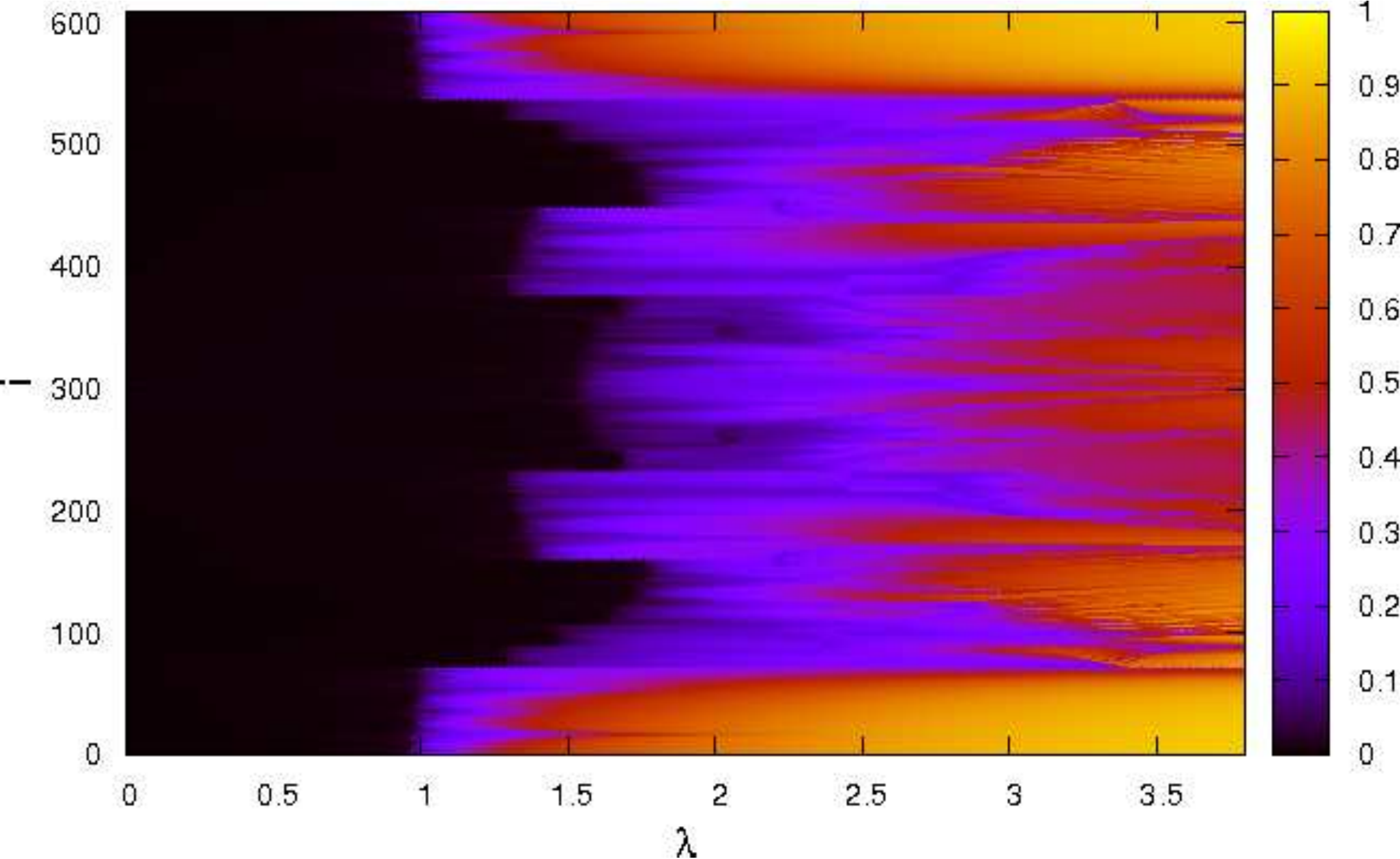}
\caption{  
The IPR values (in color) for different energy eigen-states as
a function of $\lambda$  for V = 0.5
and for a closed chain of 610 sites.
}
\label{fig2}
\end{center}
\end{figure}
\begin{figure}
\begin{center}
\includegraphics[angle=000.0,width=5in,totalheight=4.0in]{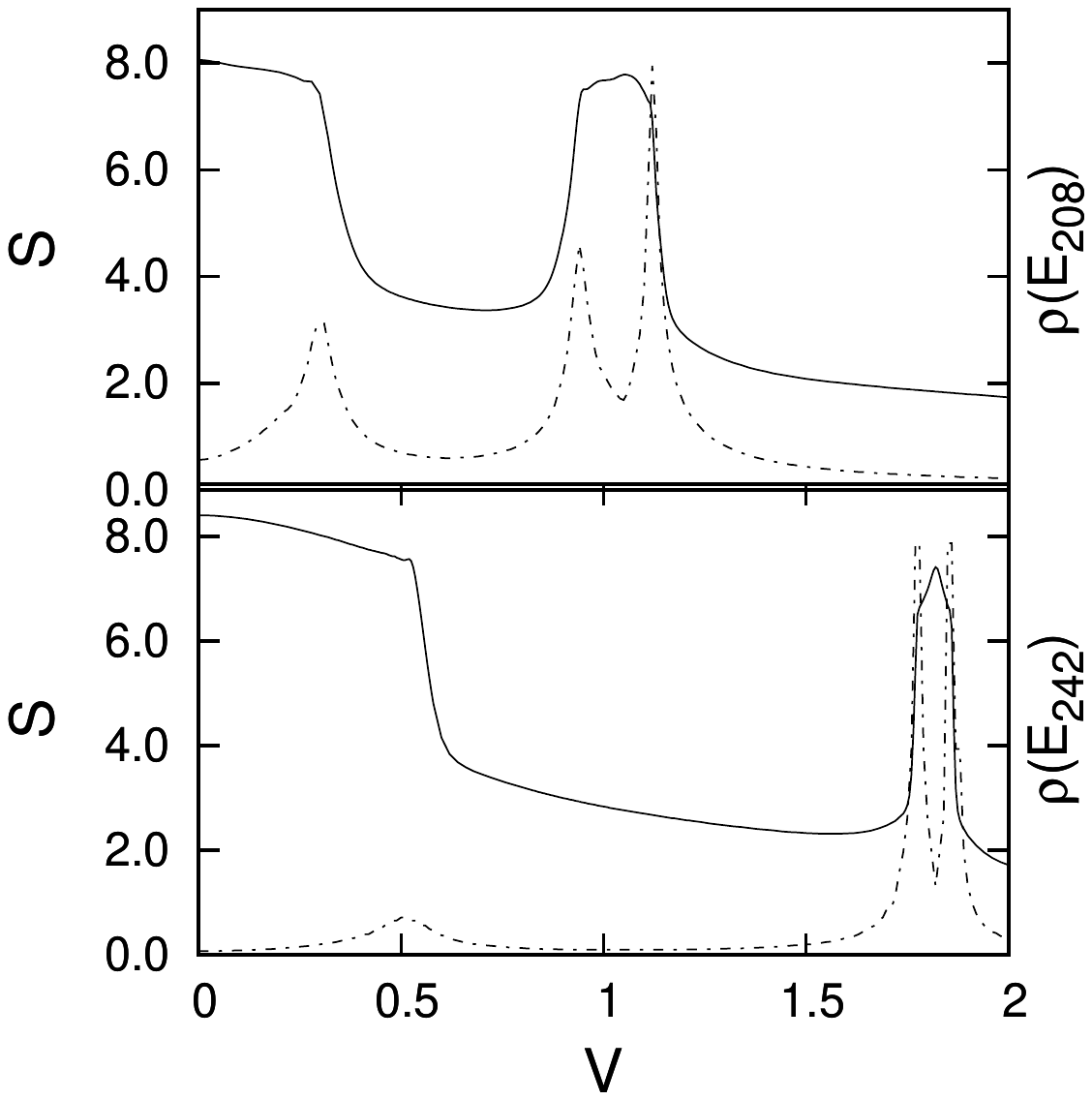}
\caption[]
{
The variation of the entanglement $S$ (solid line) with V of the 
eigen-energy levels 208 (top panel) and 242 (bottom panel) 
for $\lambda$ = 1.5. The variations  of
the density of states (DOS)($\rho(E_{208})$ and $\rho(E_{242})$)
for these eigen-energy levels are also shown (dash-dot).}
\label{fig3}
\end{center}
\end{figure}
In Fig. 1 we have displayed the IPR values of all the 610 states
for various values ($<$ 2) of $\lambda$. For these values of $\lambda$, all 
the states are extended in the absence of $V$. We find that increasing $V$
drives most states to localized. We also notice that the band edge states 
are more susceptible to localization and become localized for smaller values 
of $V$.
When we increase $\lambda$ another
qualitatively different behavior is found. In the results
shown for $\lambda$ = 1.3 and 1.5, we notice that some states
exhibit a re-entrant localization-delocalization transition.  
We do not find any re-entrant behavior with 
increasing $\lambda$ for a fixed $V$ as shown in Fig. 2. 
A consequence of the localization-delocalization transition
is the enhancement of the entanglement with increasing $V$ 
for the concerned states 
in relevant regions, as shown in Fig. 3.
The entanglement is measured by the Shannon entropy\cite{stephan} defined by
\begin{equation}
 S\,=\,-\sum_{i}p_{i}\,log_{2}\,p_{i}.
\end{equation}
The variation of the single particle spectrum with $V$ for $\lambda$ = 1.5
is shown in Fig. 4. It is clear that for several values of $V$
the system develops {\em multiple} mobility edges. 
This may be contrasted with the development of a single
mobility edge due next nearest neighbor (NNN) hopping in the 
absence of $V$\cite{boers,biddle1,biddle2,biddle3}.
The development of mobility edges due to $V$ can also be seen 
in Fig. 5, where we have plotted the single particle spectrum and the 
corresponding IPR values as a function of $\lambda$ for $V=0.5$. One also
notices that beyond $\lambda\,>\,2$, all the states are localized.
Further, while $\lambda$ shifts more levels in to the central gap
region, the $V$ has the opposite effect.
Note also the development of multiple mobility edges in
some range of values of $\lambda$. 
We also find that the delocalization to 
localization transition or vice versa for any energy eigen-state is 
associated with a rise in the single particle density of states (DOS).
The variations of the DOS with $V$ for two 
eigen-energies are also shown in Fig. 3. 
It is seen that whenever $S$ drops or rises rapidly signaling a 
delocalization-localization transition or vice versa 
the DOS shows a peak.
At this juncture we are unable to provide an analytical
explanation of these results.
Clearly, the superlattice potential
destroys the self-duality of the AA model,
and a complex interplay of the AA quasi-disorder and the superlattice
potential lead to new qualitative features like the re-entrant
localization-delocalization transition and the development of multiple
mobility edges.
Considering cold atom experiments in mind, we have
checked the effect of an overall harmonic confining potential.
In this case the system Hamiltonian becomes
\begin{equation}
\tilde{H}=H+\sum_{i}ki^{2}c^{\dag}_{i}c_{i},
\end{equation}
where $k$, which has an energy unit, is the harmonic
confining potential strength. For the harmonic trap, we
use open boundary conditions which is more appropriate.
The results of these calculations are shown in Fig. 6.
The confining potential clearly drives more states localized.
This behavior is qualitatively expected as a confining potential
generally favors localization.
\begin{figure}
\begin{center}
\includegraphics[angle=270.0,width=4in,totalheight=4.5in]{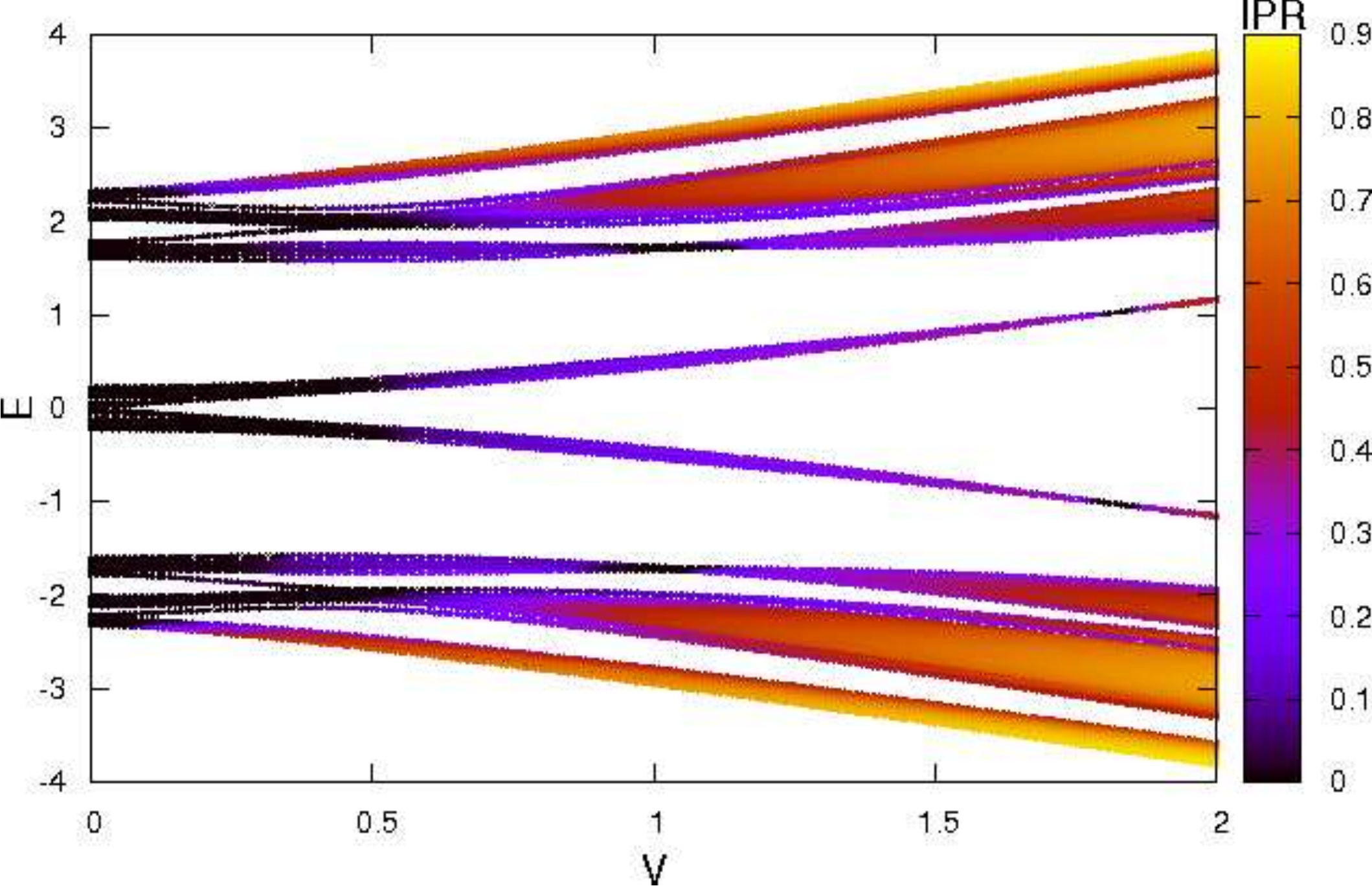}
\caption[]
{
The single particle spectrum as a function of V for 
$\lambda$ =1.5 for a closed chain of 610 sites. The IPR values of the 
corresponding states are shown in color.
}
\label{fig4}
\end{center}
\end{figure}
\begin{figure}
\begin{center}
\includegraphics[angle=270.0,width=4in,totalheight=3.5in]{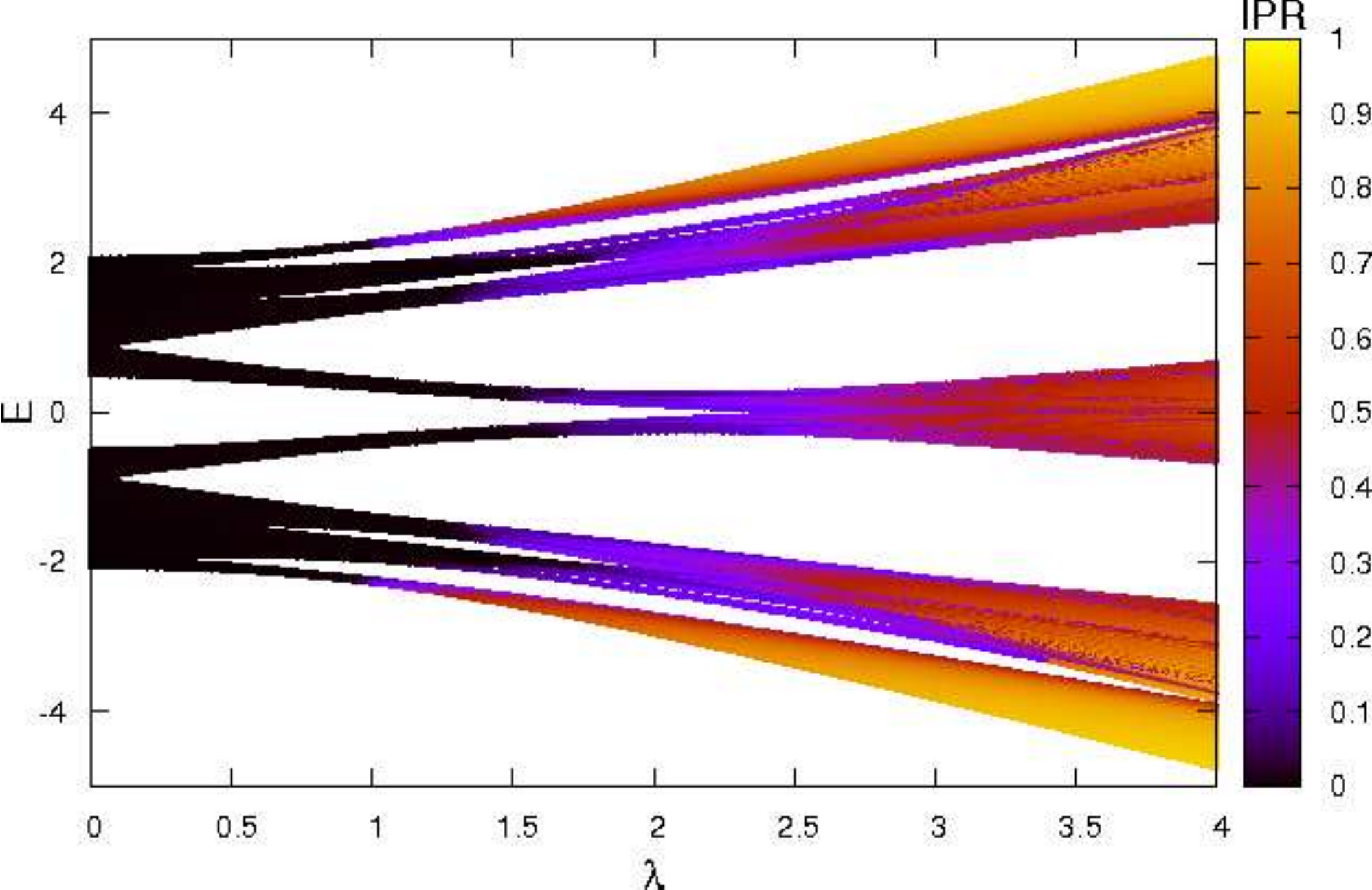}
\caption[]
{
The single particle spectrum as a function of  
$\lambda$ for V = 0.5 for a closed chain of 610 sites. The IPR
values are shown in color. 
}
\label{fig5}
\end{center}
\end{figure}
\begin{figure}
\begin{center}
\includegraphics[angle=270.0,width=4in,totalheight=3.5in]{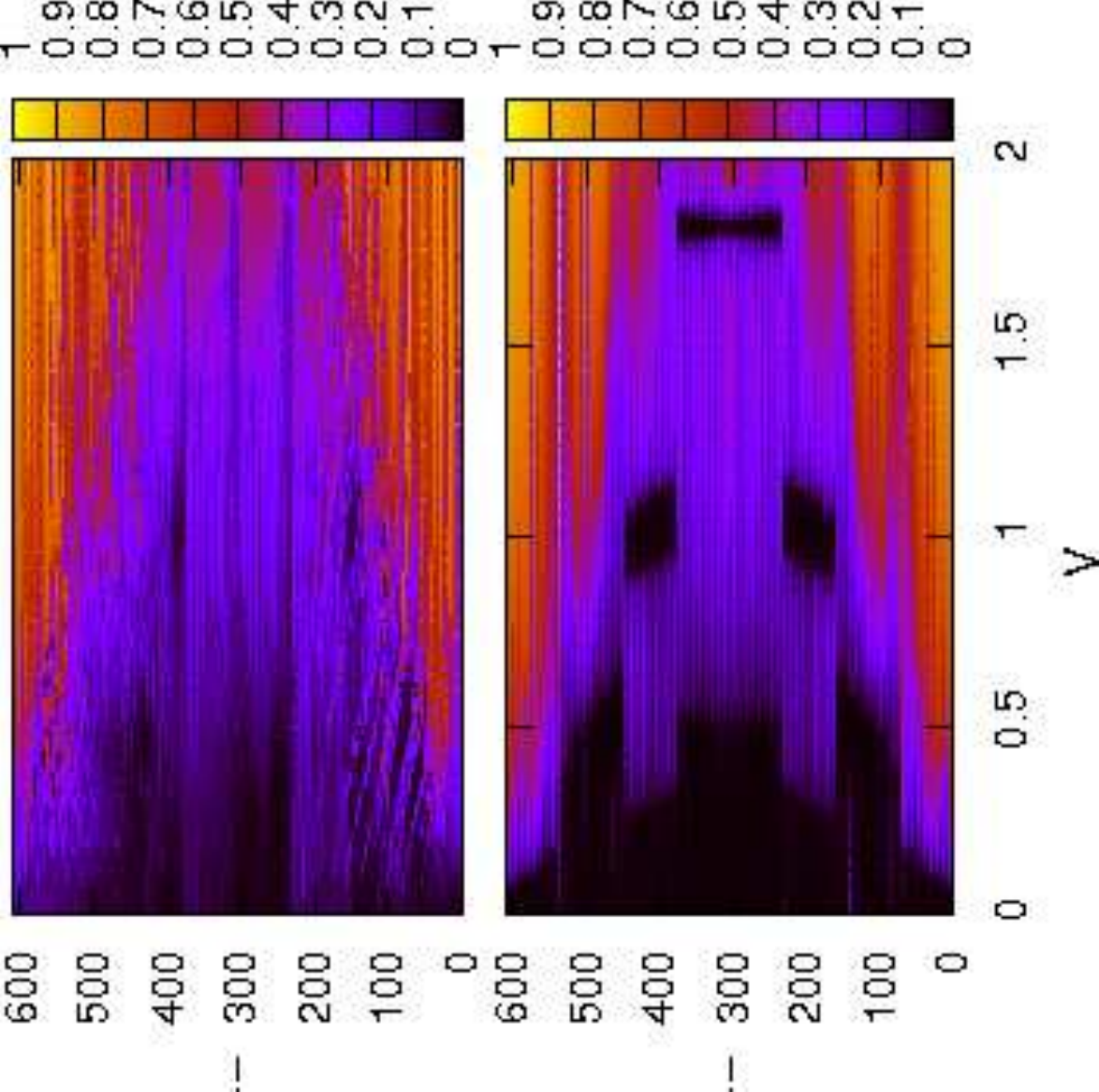}
\caption[]
{
The effect of a harmonic potential on the
IPR values (in color) for different energy eigen-states as
a function of V for $\lambda$  = 1.5
for an open chain of 610 sites.
The results are shown for k = 0.00001 (top panel) and for k = 0 (bottom panel).
The harmonic potential is placed symmetrically about the center
of the lattice.
}
\label{fig6}
\end{center}
\end{figure}
\section{The influence of the superlattice potential on 
the Bose-Einstein condensation}
\label{sec3}
In the previous section we have shown the superlattice potential
leads to significant changes in the nature of the eigen-states
of the AA model. These changes may influence the collective
properties like Bose-Einstein condensation of a many-boson
system (see also the note in Ref. \onlinecite{note}). 
In this section, we study the effect of $V$ on the
ground state occupancy of a many-boson system in AA potential
with an overall harmonic confining potential.
The Hamiltonian of this system of lattice bosons is then:
\begin{equation}
\tilde{\tilde{H}}=\tilde{H}-\mu\sum_{i}c^{\dag}_{i}c_{i},
\end{equation}
where $\mu$ is the chemical potential.
The boson number equation is
\begin{equation}
N\,=\,\sum_{i=1}^{m}N(E_{i}),
\end{equation}
where $E_i$'s are the energy levels obtained by diagonalizing
$\tilde{H}$ and
\begin{equation}
N(E_{i})=\frac{1}{e^{\beta\,(E_{i}-\mu)}-1},
\end{equation}
where $\beta\,=\,1/k_{B}T$ in which $k_{B}$ is the Boltzmann constant
and $T$ the temperature. For a fixed boson number and temperature,
the chemical potential and
then the boson populations in various energy levels are
calculated using the boson number equation. 
The temperature variation of the fractional ground state
population($N_0/N$), where $N_0\, =\, N(E_1)$ is the ground state 
population and $N$ the total number of bosons, for various values 
of $V$ and $\lambda$ = 1.5 is shown in Fig. 7. 
While in the low temperature range ($k_BT <$ 0.1), the $N_0/N$
decreases with $V$, it increases with $V$ in the higher
temperature range. The behavior of $N_0/N$ is determined
by the low energy levels. We examined the $V$ dependence
of the energies of the lowest few excited states. The results
for $\lambda=~1.5$ are shown in Fig. 8. 
\label{sec4}
\begin{figure}
\begin{center}
\includegraphics[angle=000.0,width=4in,totalheight=3in]{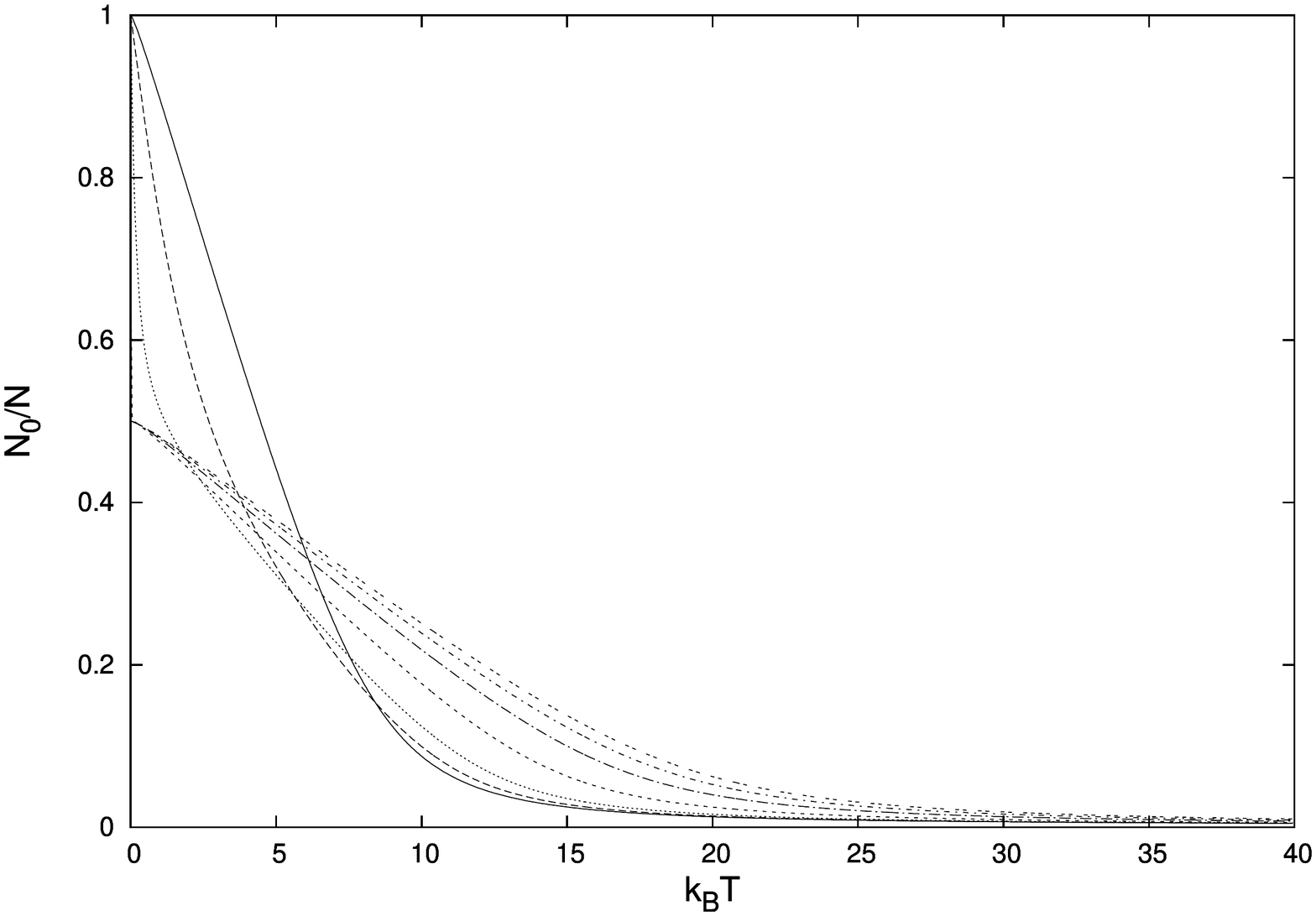}
\caption[]
{
The variation of the condensate fraction with temperature
for 10000 bosons in an open chain with 610 sites in
a harmonic trap with $k$ = 0.00001 
and $\lambda$ = 1.5. The results are for:
V = 0 (solid line), 0.1 (long dash), 0.2 (dot), 
0.5 (short dash), 1.0 (dash-dot),
1.5 (short dash-dot), and 2.0 (double dashes). 
}
\label{fig7}
\end{center}
\end{figure}
\begin{figure}
\begin{center}
\includegraphics[angle=000.0,width=4in,totalheight=2.5in]{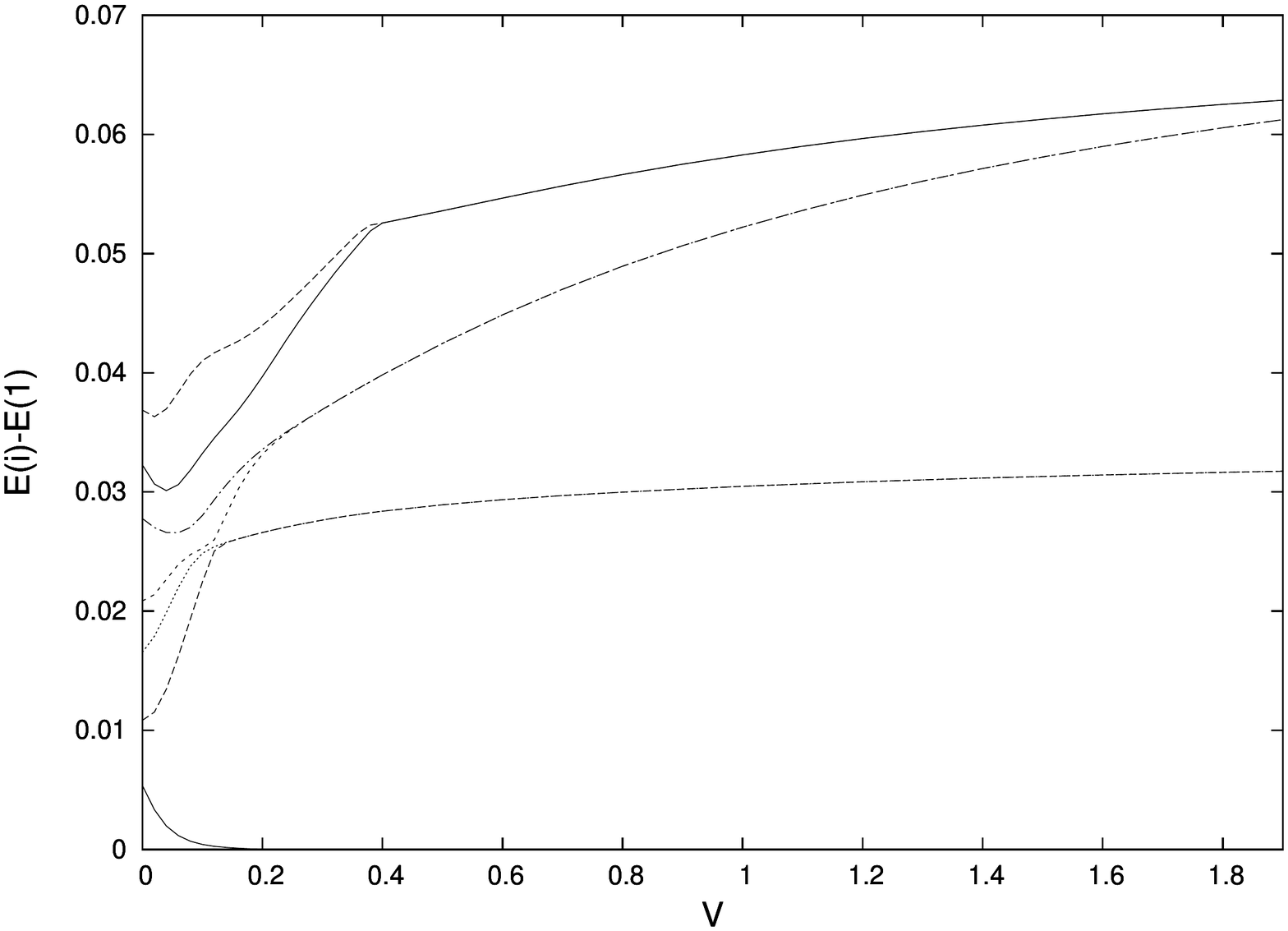}
\caption[]
{
The variation of the difference between energies of 
the ground state  and first few excited states $E(i)-E(1)$
for (from bottom to top) i = 2, 3, 4, 5, 6, 7, and 8, for
$\lambda\,=\,1.5$.
}
\label{fig8}
\end{center}
\end{figure}
The first excited state move close to the
ground state with increasing $V$ leading to a rapid thermal
depletion of the ground state. Beyond $V >$ 0.2, the ground
state and the first excited state
become almost degenerate and the bosons are
distributed equally in these levels 
leading to a drop of $N_0/N$ from 1 to about 0.5
as seen in Fig. 7.
We also see that the second excited state onwards, the
excited state energies move away from the ground state
with increasing $V$. So, in the higher
temperature range, the population of these excited states
decrease with increasing $V$ leading to an increase in $N_0/N$.

\section{Conclusions}
In this paper, we first investigated the effect
of a superlattice potential on the single particle localization
properties of a lattice boson in the presence of AA disorder.
We find a re-entrant localization-delocalization
transition and the development of multiple mobility edges
for range of superlattice potential strengths.
We also studied the Bose-Einstein
condensation with an additional harmonic trapping potential.
It is found that while an increase in the superlattice potential leads to
an increase in the depletion of the condensate in the low
temperature limit, it has the opposite effect beyond this 
temperature regime. These trends of the condensate fraction
was argued to be resulting from the changes in the low 
energy single particle spectrum produced by the superlattice
and the quasi-disorder potentials.
\begin{acknowledgments}
RRK thanks Professor B. K. Chakrabarti, Director, SINP and
Professor R. Ranganathan, Head, CMP Division, SINP
for hospitality at SINP.
\end{acknowledgments} 

\end{document}